# Superconductivity at 11.3 K induced by cobalt doping in CeOFeAs

J. Prakash[a], S. J. Singh[b], S. Patnaik[b] and A. K. Ganguli[a]*

[a]Department of Chemistry, Indian Institute of Technology, New Delhi 110016 India

[b]School of Physical Sciences, Jawaharlal Nehru University, New Delhi 110067 India

## Abstract

Pure phases of a new oxyarsenide superconductor of the nominal composition $CeOFe_{0.9}Co_{0.1}As$ was successfully synthesized by solid state reaction in sealed silica ampoules at 1180 C. It crystallizes in the layered tetragonal ZrCuSiAs type structure (sp gp P4/nmm) with lattice parameter of '**a**' = 3.9918(5) Å and '**c**' = 8.603(1) Å. A sharp superconducting transition is observed at 11.31 K (± 0.05 K) with an upper critical field of 45.22 T at ambient pressure. The superconducting transition temperature is drastically lowered ( ~ 4.5, 4.9 K) on increasing the concentration (x = 0.15, 0.2) of cobalt.

***Author for correspondence**

Email:- ashok@chemistry.iitd.ernet.in

Tel No. 91-11-26591511

Fax 91-11-26854715

## Introduction

Superconductivity in the oxypnictide $LaO_{1-x}F_xFeAs$ with a transition temperature ($T_c$) of 26 K [1] has rejuvenated the search for new superconductors. $T_c$ was enhanced to 43 K at high pressure [2] and it was also shown that La can be replaced with rare earth ions of smaller ionic radii like Sm, Ce, Nd, Pr and Gd [7]. The highest $T_c$ achieved so far in these oxypnictides is 56 Kin $Gd_{1-x}Th_xFeAs$ [8].

LnOFeAs type compounds (Ln= rare earth) [9] crystallize in the layered tetragonal (ZrCuSiAs) structure (Space group: P4/nmm) which can be described as a filled variant of the PbFCl structure type which consist of two dimensional layers of edge-sharing $FeAs_{4/4}$ tetrahedra alternate with sheets of edge-sharing $OLn_{4/4}$ tetrahedra. The ionic Ln-O layers act as a charge reservoir and metallic FeAs layers act as charge carriers [2]. The parent CeOFeAs is semimetal in nature (not superconducting) with anomalies in the electrical conductivity and magnetic susceptibility plots at about 155 K [10] due to an unstable spin density wave (SDW) and distortion of structure below this temperature. This distortion can be suppressed leading to superconductivity by either doping $F^-$ in place of $O^{2-}$ which enhances the charge carriers in the FeAs layers or with hole doping which can be achieved by doping lower-valent ions like $Sr^{2+}$ at Ln [11] site. We have earlier shown that KF doping in LaOFeAs also leads to superconductivity [12]. The deficiency of oxygen in LnOFeAs (Ln= rare earth metals) also leads to superconductivity with maximum Tc of 55 K in $SmO_{0.85}FeAs$ [13]. Recently it has been shown that superconductivity can be induced in LaOFeAs and SmOFeAs by cobalt substitution [14-16] in place of iron which results in direct injection of electrons in the conducting FeAs layers.

We have investigated the effect of substitution of cobalt in the cerium oxyarsenide (CeOFeAs) to study the effect of variation of the transition metal and to study the variation of the rare-earth in this oxypnictides. We find that for low doping (x=0.1) a $T_c$ of 11.3 K is obtained while for higher x-values a metallic behaviour is observed. We have carried out field dependent studies and evaluated the upper critical field for thisnew superconducting compound.

## Experimental

For the synthesis of $CeOFe_{1-x}Co_xAs$ (x= 0.1 – 0.2), stoichiometric amounts of Ce, As, $CeO_2$, $Co_3O_4$ and FeAs were sealed in evacuated silica ampoules ($10^{-4}$ torr) and heated at 950 C for 24h . The powder was compacted (5 tonnes) and the disks were wrapped in Ta foil, sealed in evacuated silica ampoules and heated at 1180 C for 48 h. All chemicals were handled in a nitrogen filled glove box. Powder X-ray diffraction patterns of the finely ground powders were recorded with Cu-Kα radiation in the 2θ range of 20° to 70°. The lattice parameters were obtained from a least squares fit to the observed *d* values.

The resistivity measurements were carried out using a cryogenic 8 T cryogen-free magnet in conjunction with a variable temperature insert (VTI). In this system the samples were cooled in helium vapor. Standard four probe technique was used for transport measurements. Contacts were made using 44 gauge copper wires with air drying conducting silver paste. The external magnetic field (0-5 T) was applied perpendicular to the probe current direction and the data were recorded during the warming cycle with heating rate of 1 K/min. The real part of the magnetic susceptibility was measured as discussed earlier [17]. A change in magnetic state of the sample results in a change in the inductance of the coil and is reflected as a shift in the oscillator frequency which is measured by an Agilent 53131A counter.

## Results and Discussion

Figure 1 shows the powder X-ray diffraction patterns of the compounds with nominal composition $CeOFe_{1-x}Co_xAs(0.1\leq x\leq 0.2)$. All the observed reflections could be satisfactorily indexed based on the tetragonal CeOFeAs(space group P4/nmm) phase. The refined lattice parameters show a decrease with increase in Co- substitution (Fig. 2) which is expected since the ionic size of $Co^{2+}$ is smaller than that of $Fe^{2+}$. This indicates the substitution of cobalt in place of iron in CeOFeAs. Note that all the three phases are pure unlike the earlier reports on Co-doped SmOFeAs (where unidentified impurity phases are reported) [16] and Co-doped LaOFeAs (where $La_2O_3$ and FeAs are reported) [14]

The zero field resistivity as a function of temperature for $CeOFe_{0.9}Co_{0.1}As$ is shown in Figure 2. A metallic behaviour was observed down to 90 K and then a semimetallic / semiconducting behavior exists between 90 K and ~ 12 K(Inset of Fig. 2) before a sharp superconducting transition at 11.3 K(Fig.2) with a width of 1.3 K. This is much sharper than those reported for the La and Sm analogues. The magnetic studies however show the onset of diamagnetism at ~ 10.0 K (inset of Fig.2). On higher cobalt doping ( x= 0.15 and 0.2) we find nearly metallic behavior down to ~ 5 K below which superconductivity is observed (Fig.3). La and Sm analogues with x = 0.15 composition also show superconductivity at lower temperatures than the x = 0.1 composition. It thus appears that optimal doping of electrons (highest $T_c$) for all three systems (La, Ce and Sm) occurs at x = 0.1.

The in - field resistivity transitions for the sample $CeOFe_{0.9}Co_{0.1}As$ are shown in Figure 5. At an applied field of 5 Tesla the offset of transition shifts by ~ 3.5 K .The inset of Figure 3 shows the upper critical field ($H_{c2}$) and irreversibility field (H*) as a function of temperature obtained from magnetic field-dependent resistivity studies. We have used the 90% of normal state resistivity $\rho_n$ (at T = $T_c$) criteria to define $Hc_2$ [18] and 10% criteria for the corresponding H*. The slope $dH_{c2}/dT$ is estimated to be – 5.8 T/K. The extrapolated $H_{c2}$ (0) values using the Werthamer-Halfand-Hohenberg (WHH) formula $H_{c2}(0) = -0.693\ T_c\ (dH_{c2}/dT)$ gives an estimate of zero temperature upper critical field ($H_{c2}$) of 45.22 T.

In conclusion, we have succeeded in obtaining pure phases of new superconductors by doping a magnetic ion (Co) in the FeAs planes of the non-superconducting cerium oxypnictides (CeOFeAs). This indicates a major difference from cuprate superconductors where any substitution in the conducting $CuO_2$ planes by other metal ions always destroys superconductivity or lowers the transition temperature.

**Acknowledgement**

AKG and SP thank DST, Govt. of India financial support. JP and SJS thank CSIR and UGC, Govt. of India, respectively for fellowships.

**Figure captions**

**Figure 1.** Powder X-ray diffraction patterns (XRD) pattern of composition $CeOFe_{1-x}Co_xAs$ (a) x=0.1 (b) x= 0.15 and (c) x=0.20 annealed at 1180 C.

**Figure 2.** Lattice parameters as a function of Cobalt concentration

**Figure 3.** Temperature dependence of resistivity (ρ) as a function of temperature for the composition $CeOFe_{0.9}Co_{0.1}As$ annealed at 1180 C. Inset show resistivity till higher temperatures and the inductive part of susceptibility as a function of temperature.

**Figure 4.** Temperature dependence of resistivity (ρ) as a function of temperature for the composition $CeOFe_{1-x}Co_xAs$(■) x= 0.15 and (●) x=0.20 annealed at 1180 C.

**Figure 5.** Temperature dependence of resistivity (ρ) in the presence of magnetic field (0 T, 1 T. 3 T and 5 T). Inset shows the upper critical field ($H_{c2}$) and irreversibility field (H*).

**Figure 1**

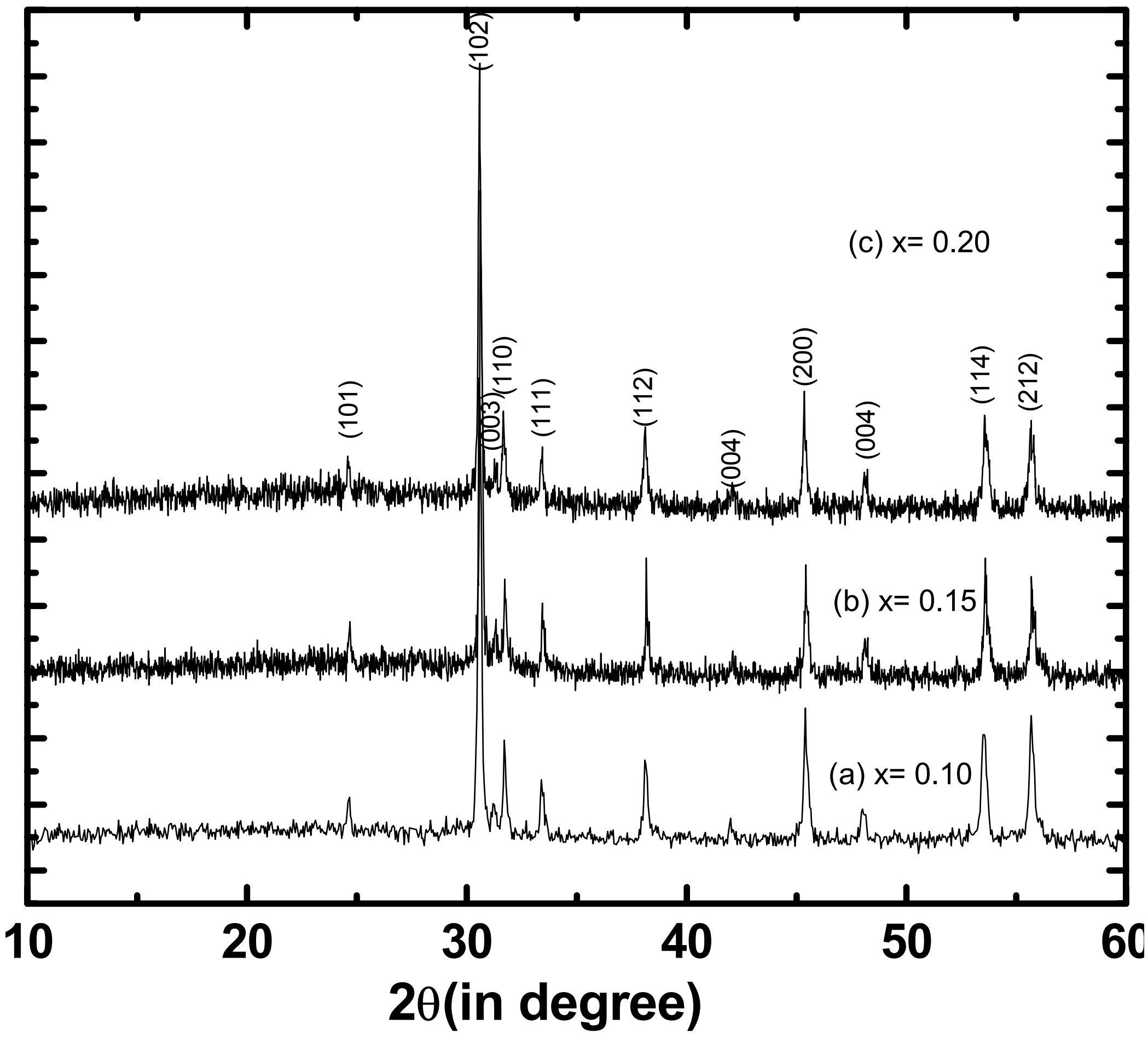

**Figure 2**

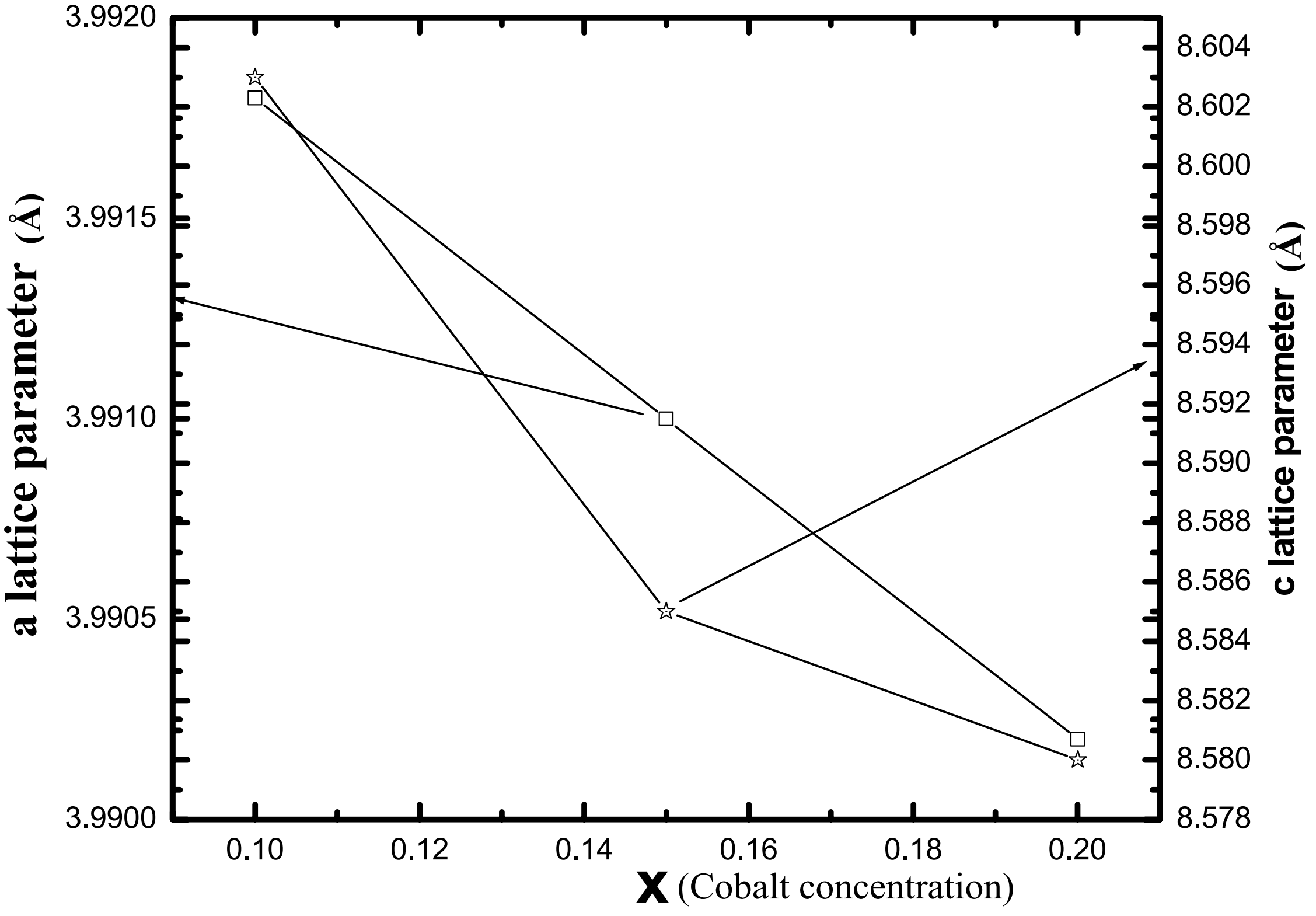

**Figure 3**

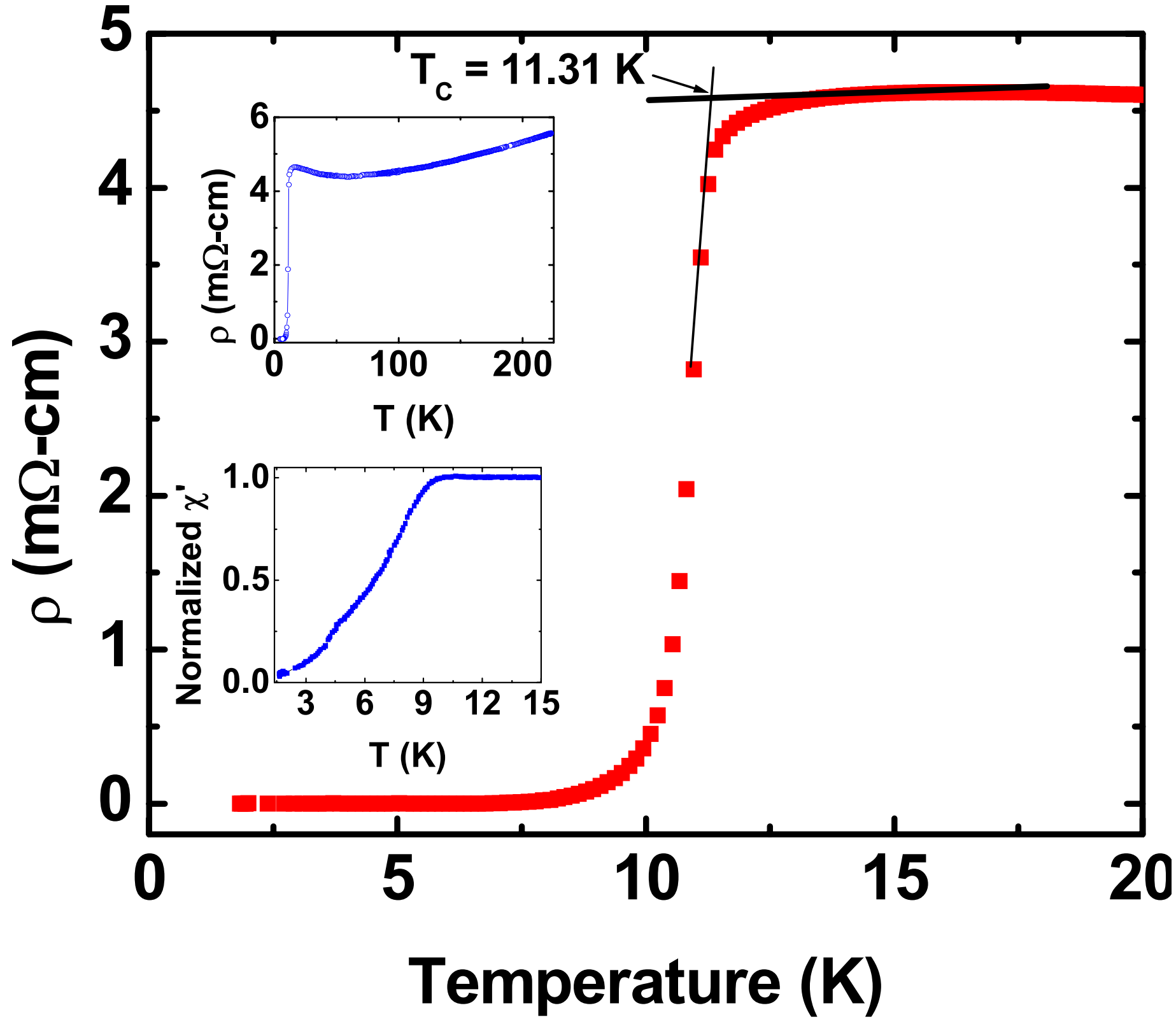

Figure 4

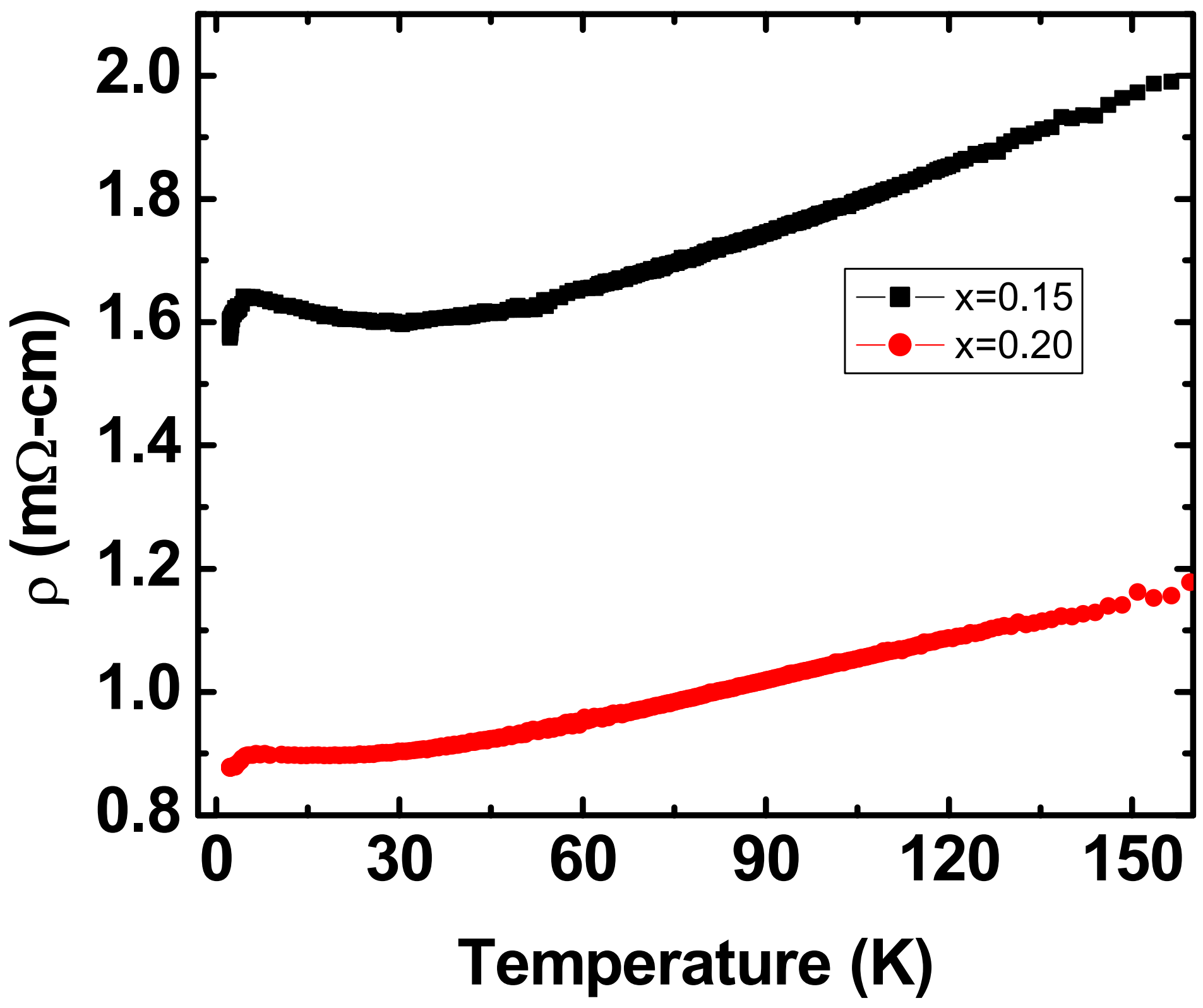

**Figure 5**

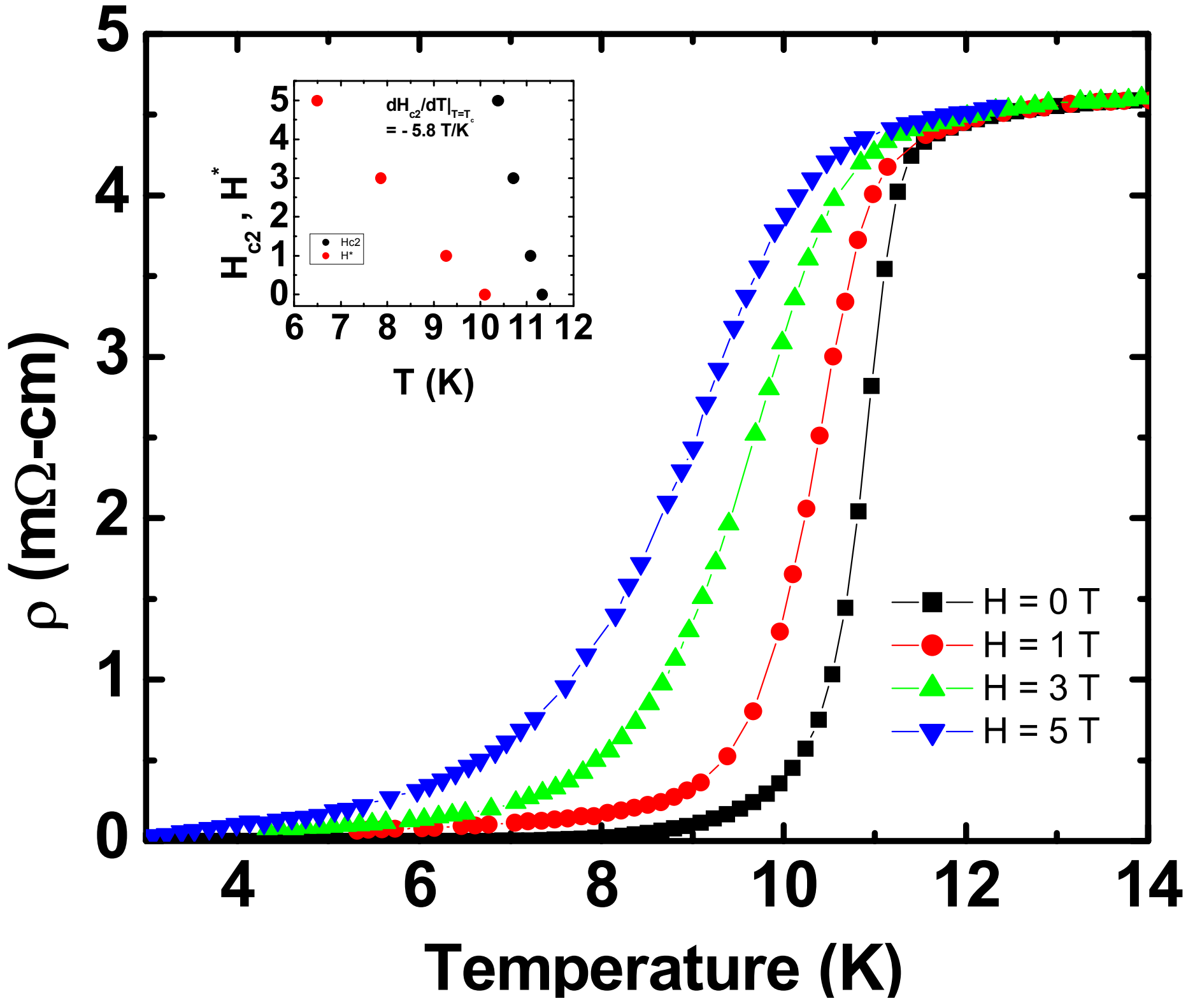

ρ (mΩ-cm)
Temperature (K)
H = 0 T
H = 1 T
H = 3 T
H = 5 T
dH_c2/dT|_T=Tc
= - 5.8 T/K
Hc2
H*
H_c2 , H*
T (K)